\documentclass[article]{./aa}
\usepackage{natbib}
\usepackage{graphicx}
\usepackage{sidecap}
\usepackage{color}
\begin{document}
   \title{Magnetic field intensification: comparison of 3D MHD simulations with Hinode/SP results}
   \titlerunning{Magnetic field intensification: 3D MHD vs. Hinode/SP}
   \author{S. Danilovic \inst{1} \and M. Sch{\"u}ssler \inst{1} \and
           S.K. Solanki \inst{1,2}}

   \institute{Max-Planck-Institut f\"ur Sonnensystemforschung,
              Max-Planck-Stra{\ss}e 2, 37191 Katlenburg-Lindau,
              Germany \and
              School of Space Research, Kyung Hee University,
              Yongin, Gyeonggi, 446-701, Korea}

   \date{\today}

  \abstract
  {Recent spectro-polarimetric observations have provided
  detailed measurements of magnetic field, velocity and intensity during
  events of magnetic field intensification in the solar photosphere.}
  {By comparing with synthetic observations derived from MHD
    simulations, we aim to discern the physical processes underlying the
    observations, as well as to verify the simulations and the
    interpretation of the observations.}
  {We consider the temporal evolution of the relevant physical
  quantities for three cases of magnetic field intensification in a
  numerical simulation. In order to compare with observations, we
  calculate Stokes profiles and take into account the spectral and spatial
  resolution of the spectropolarimeter (SP) on board Hinode. We determine
  the evolution of the intensity, magnetic flux density and
  zero-crossing velocity derived from the synthetic Stokes parameters,
  using the same methods as applied to the Hinode/SP observations to
  derive magnetic field and velocity information from the
  spectro-polarimetric data. }
  {The three events considered show a similar evolution:
  advection of magnetic flux to a granular vertex, development of a
  strong downflow, evacuation of the magnetic feature, increase of the
  field strength and the appearance of the bright point. The magnetic
  features formed have diameters of 0.1-0.2\arcsec. The downflow
  velocities reach maximum values of 5-10 km/s at $\tau=1$.  In the
  largest feature, the downflow reaches supersonic speed in the lower
  photosphere. In the same case, a supersonic upflow develops
  approximately $200$~s after the formation of the flux
  concentration. We find that synthetic and real observations are
  qualitatively consistent and, for one of the cases considered, agree
  very well also quantitatively. The effect of finite resolution (spatial smearing) is
  most pronounced in the case of small features, for which the synthetic
  Hinode/SP observations miss the bright point formation and also the high-velocity downflows during the formation of the smaller magnetic features.}
  {The observed events are consistent with
  the process of field intensification by flux advection, radiative
  cooling, and evacuation by strong downflow found in MHD simulations.
  The quantitative agreement of synthetic and real observations
  indicates the validity of both the simulations and the interpretations
  of the spectro-polarimetric observations.}

 \maketitle

\section{Introduction}

Magnetic field is ubiquitously present in the solar photosphere
\citep{deWijn:etal:2008}. On granular scales, it undergoes
continual deformation and displacement. It is swept by the
horizontal flows and concentrated in the intergranular lanes.
Flows are able to compress the field so that the magnetic energy
density $B^{2}/8\pi$ approaches the kinetic energy density $\rho
v^{2}/2$ of the flow \citep{Parker:1963,Weiss:1966}. This results
in a magnetic field strength of a few hundred Gauss at the solar
surface. Further intensification to kG strength is driven by the
mechanism referred to as: \textit{superadiabatic effect}
\citep{Parker:1978}, \textit{convective collapse}
\citep{Webb:Roberts:1978,Spruit:Zweibel:1979} or
\textit{convective intensification}
\citep{GrossmannDoerth:etal:1998}.

The first two concepts are a theoretical idealization of the
process. The superadiabatic effect contains the basic idea.
\citet{Parker:1978} pointed out that a thermally isolated
dowflowing gas within the flux tube in a superadiabatically
stratified environment will be accelerated, which would lead to
evacuation of the flux tube. Because of the resulting pressure
deficit, the gas inside the flux tube will then be pressed
together (together with the frozen-in magnetic field) by the
surrounding gas, causing the magnetic pressure to increase until a
balance of total pressure (magnetic + gas) is reached.

The convective collapse extents the concept to the convective
instability. It starts with a flux tube in thermal and mechanical
equilibrium with the surrounding hydrostatically superadiabaticly
stratified plasma. Since external stratification is convectively
unstable, any vertical motion within the flux tube can be
amplified. Downward flow will grow in amplitude and drain the
material from the flux tube. The process continues until a new
equilibrium with a strong field is reached. Different aspects of
the concept have been the subject of extensive research
\citep[see][for reviews]{Schuessler:1990,Steiner:1999}.
\begin{figure*}
    \centering
    \includegraphics[width=0.95\hsize,]{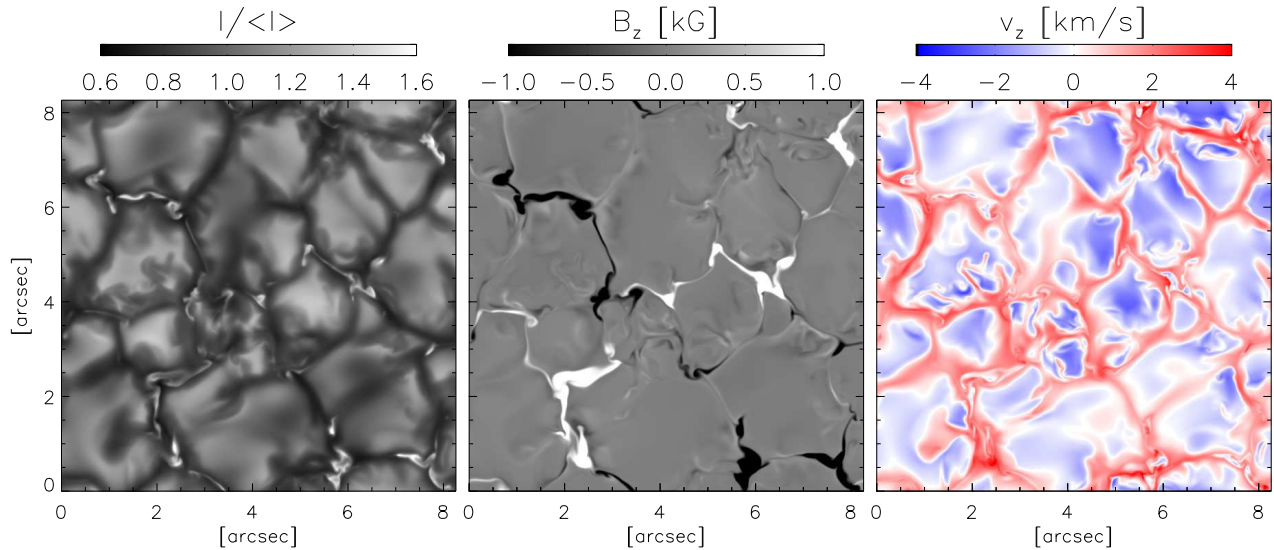}
    \caption{Maps of the whole simulation domain at $t=140$~s.
    Normalized continuum intensity at $630$~nm(left), vertical component of magnetic field (middle) and velocity (right) at a geometrical height roughly corresponding
     to the level of $\langle\tau_{500}\rangle = 1$ ($\approx 930$~km above the bottom of the computational box).}
    \label{fig:snap}
\end{figure*}
The term convective intensification is used for magnetic field
intensification in realistic MHD simulations, where the process
occurs in its full complexity. It is driven by the thermal effect
in the surface layer of the magnetic concentration. There, due to
the presence of magnetic field, heat transport by convection is
reduced. The material inside the concentration radiates more that
it receives. This leads to cooling of material which thus starts
to sink and partial evacuation of the concentration occurs.
Contraction of the magnetic concentration by the surroundings
(result of the pressure imbalance) leads to an increase in
magnetic field strength. Thus, the simulations
\citep{Nordlund:1983,GrossmannDoerth:etal:1998,Gadun:etal:2001,Voegler:etal:2005,Cheung:etal:2008}
bear out the basic properties described by idealized concepts.
That is the downflow, the evacuation of the magnetic structure,
the field increase and, in some cases, establishment of a new
equilibrium. The 3D MHD simulations show that the strong magnetic
concentrations form as the horizontal flows in the intergranular
lanes advect weak, nearly vertical field and concentrate it at the
vertices of granular and mesogranular downflow lanes
\citep{Stein:Nordlund:1998,Stein:Nordlund:2006}. Larger magnetic
structures form at sites where a granule submerges and the
surrounding field is pushed into the resulting dark region.
Whether the formed concentration appears dark or bright in the
continuum intensity depends on whether the vertical cooling is
compensated or not by the lateral heating due to horizontal energy
exchange \citep{Bercik:etal:2003,Voegler:etal:2005}. This
formation scenario is consistent with the observations described
by \citet{Muller:1983} and \citet{Muller:Roudier:1992}. Their
observations show that network bright points form in intergranular
spaces, at the junction of converging granules as the magnetic
field gets compressed by the converging granular flow.

The 2D simulations by \citet{GrossmannDoerth:etal:1998} revealed
that magnetic flux concentrations formed by convective
intensification can evolve in different ways. They present two
possible outcomes. Depending on the initial magnetic flux, a
magnetic concentration can reach a stable state after the process,
or can be dispersed due to an upflow that develops as high speed
downflowing material rebounces from the dense bottom of the tube.
Similar results were presented by \citet{Takeuchi:1999} and
\citet{Sheminova:Gadun:2000}.

Observational evidence was found for both cases.
\citet{BelloGonzalez:etal:2008} reported on the formation of a
magnetic feature at the junction of intergranular lanes, without
any significant upflow observed. \citet{BellotRubio:etal:2001}, on
the other hand, detected a strongly blueshifted Stokes V profile
originating in a upward propagating shock, $13$ minutes after the
amplification of magnetic field.
\citet{SocasNavarro:MansoSainz:2005} found that supersonic upflows
are actually quite common.

Events that are interpreted as convective collapse were detected
also with the spectropolarimeter (SP) \citep{Lites:etal:2001} of
the Solar Optical telescope \citep{Tsuneta:etal:2008} on board
Hinode \citep{Kosugi:etal:2007}. Both, \citet{Shimizu:etal:2008}
and \citet{Nagata:etal:2008} show cases of high speed downflows
followed by magnetic field intensification and bright point
appearance. The event described by \citet{Nagata:etal:2008} shows
stronger field strength and upflow in the final phase of
evolution.

In this paper we give three examples of magnetic field
intensification from MURaM simulations and make detailed
comparison with the results of \citet{Nagata:etal:2008} and
\citet{Shimizu:etal:2008}.

\begin{SCfigure*}
    \centering
    \includegraphics[width=0.75\textwidth]{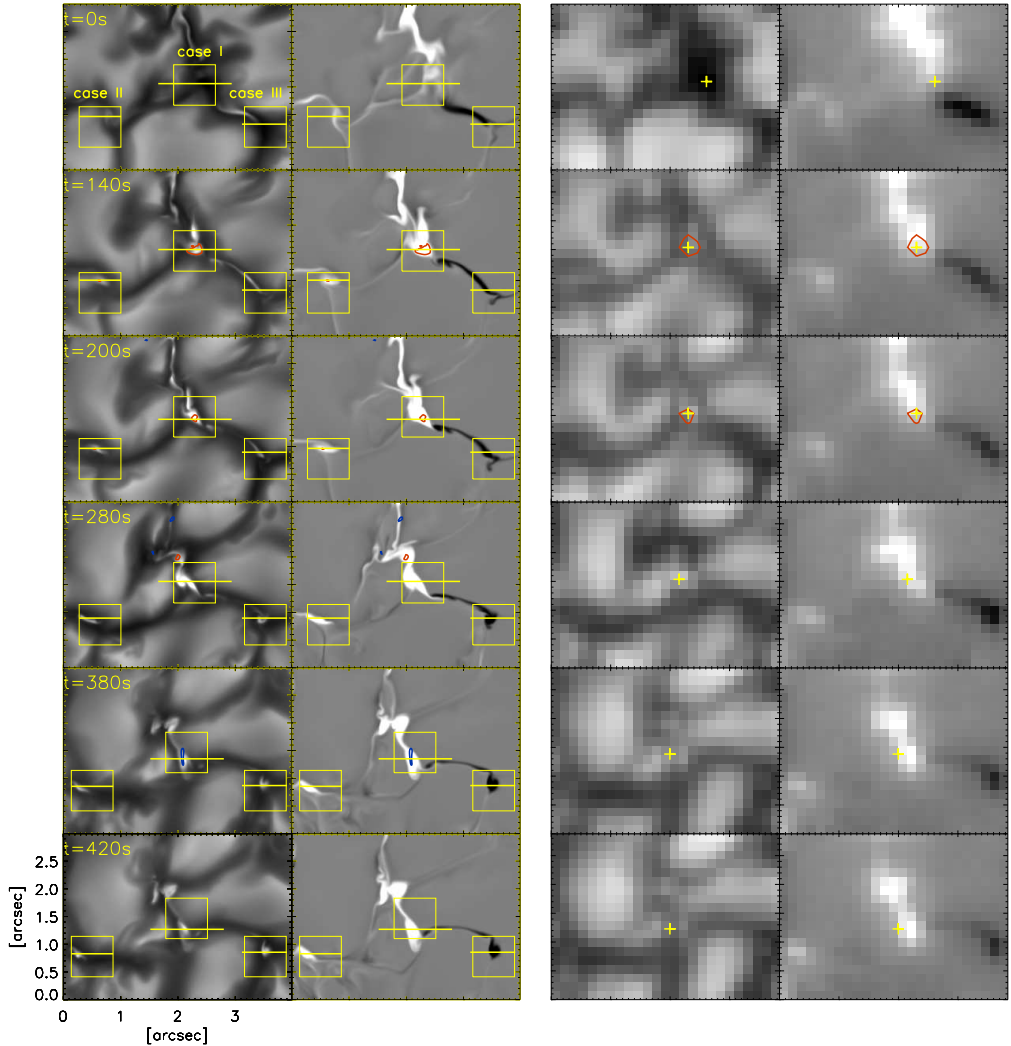}
    \caption{Evolution of the continuum intensity at $630$~nm and the magnetic field in a $4\arcsec\times
    3\arcsec$ sized region. \textit{Left-hand side} (double column): original spatial resolution; red and blue contours outline the
    downflow of 6~km/s at 80~km and upflow of 4~km/s at 400~km above $\tau = 1$, respectively;
    horizontal lines mark the positions of vertical cuts shown in Figs.~\ref{fig:cut} and ~\ref{fig:cut_f23}; boxes corresponding to cases I, II and III at
    coordinates $[2\arcsec,1.5\arcsec]$, $[0.5\arcsec,1\arcsec]$ and $[3.5\arcsec,1\arcsec]$, respectively, enlarged in Fig.~\ref{fig:evol_zoom}.
    \textit{Right-hand side}: synthetic Hinode/SP observations; left column shows continuum intensity while
    the right column shows the apparent field strength (see the text); red contour marks the region
    with $0.01$~pm of signal excess (see text); yellow crosses mark the positions of pixels that we studied in detail in
    Fig.~\ref{fig:evol_red}. Grey scales cover the range of 0.6-1.6 and 0.8-1.2 for normalized intensity
    at original and reduced resolution, respectively, $\pm 1000$~G for the vertical component of magnetic field (original resolution) and
    $\pm200$~Mx/cm$^{2}$  for the apparent longitudinal magnetic flux density (Hinode resolution).}
    \label{fig:evol}
\end{SCfigure*}

\section{Simulation data and spectral synthesis}
\label{sec:data}

We use 3D radiative MHD simulations of a thin layer containing the
solar surface carried out with the MURaM code \citep{Voegler:2003,
Voegler:etal:2005} in a $6\times6\times1.68$~Mm domain with
non-grey radiative transfer included. Vertical and horizontal
spatial resolution is 10 km and 14 km respectively. The bottom and
top boundaries are open, permitting free in and outflow of matter.
The initial magnetic field of $\langle|B|\rangle = 200$~G, is
introduced in a checkerboard-like $2\times2$ pattern, with
opposite polarity in adjacent parts. As the field is redistributed
by convective motions, opposite polarities are pushed together and
dissipated, so that the unsigned magnetic flux decreases with
time. In this way, the run simulates the decay of the magnetic
field in a mixed polarity region. Local dynamo action
\citep{Voegler:2007} does not occur since the magnetic Reynolds
number is below the threshold for dynamo action. We do not expect
that the intensification process would be significantly different
at higher Reynolds numbers (smaller grid cells). The simulated
magnetic flux concentrations considered here are well resolved.
The main effect of an increased resolution would be a decrease in
the width of the boundary layers between the flux concentration
and the surrounding downflows.

We examine a 30 min sequence of simulation snapshots with a
cadence of approximately 90 s. The snapshots shown in this paper
have $\langle|B|\rangle~\approx~150$~G at $\tau =1$. In
Fig.~\ref{fig:snap}, we show the continuum image and maps of the
vertical components of the magnetic field and velocity for a
snapshot at time $t=140$~s ($t=0$~s corresponds to the first
snapshot considered in this paper).

In order to synthesize the Stokes profiles, the physical
parameters from the simulation are used as an input for the 1D LTE
radiative transfer code, SPINOR \citep{Frutiger:etal:2000}. A
spectral range that contains Fe I lines 630.15 and 630.25 nm is
sampled in steps of $7.5$ m\AA. The Fe abundance used for the
synthesis has been taken from \citet{Thevenin:1989} and the values
of the oscillator strengths from the VALD database
\citep{Piskunov:etal:1995}. Before comparing with Hinode/SP
\citep{Lites:etal:2001} observations, the synthetic line profiles
have been treated to bring them to the same resolution as Hinode
data. Firstly, a realistic point spread function PSF
\citep{Danilovic:etal:2008} has been applied to the synthesized
Stokes profiles. The PSF takes into account the basic optical
properties of the Hinode SOT/SP system and a slight defocus which
brings the continuum contrast of the simulation to the observed
value of 7.5\%. Secondly, to take into account spectral resolution
of the spectropolarimeter, the profiles are convolved with a
Gaussian function of $25$~m\AA~FWHM and resampled to a wavelength
spacing of $21.5$ m\AA. Thirdly, noise of $10^{-3} I_{c}$ is
added. Velocities are determined from the shift between the Stokes
$V$ zero-crossing wavelength and the line core position of the
Stokes $I$ profile averaged over pixels with polarization signal
amplitudes less than $10^{-3} I_{c}$. We correct these velocities
for convective blueshift by subtracting $150$~m~s$^{-1}$. Finally,
the procedure by \citet{Lites:etal:2008} is used to calculate the
longitudinal apparent magnetic flux density (shown in
Fig.~\ref{fig:evol}).

\begin{figure}
    \centering
    \includegraphics[width=0.95\hsize,angle=0]{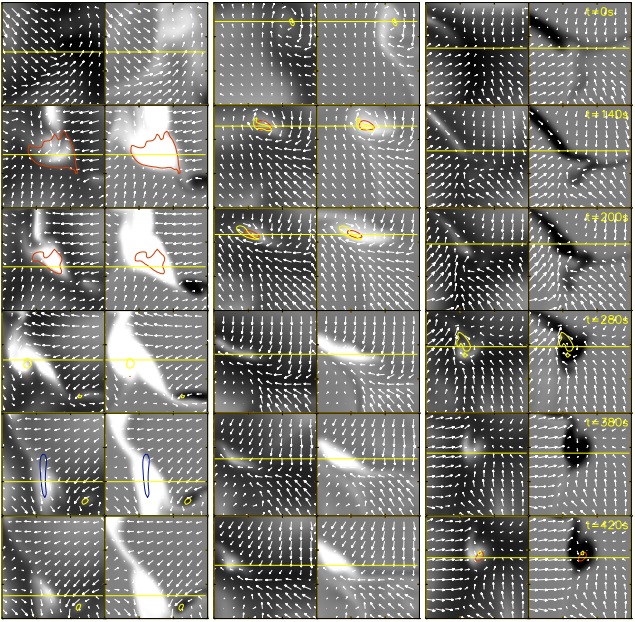}
    \caption{Enlargement of the regions, having a size of $0.75\arcsec\times 0.75\arcsec$, outlined by yellow squares in Fig.~\ref{fig:evol}. From left to
right: cases I, II and III. Continuum images (left column for each
case) and maps of the vertical components of the magnetic field
(right)  with overplotted horizontal velocities at approximately
80 km above $\tau = 1$ (white arrows). Red contours outline
locations of dowflows of 5 km/s or larger and yellow contours the
vertical component of vorticity exceeding $\pm0.2s^{-1}$ (cases I
and II) or $\pm0.1s^{-1}$ (case III).}
    \label{fig:evol_zoom}
\end{figure}

\section{Results}

Fig.~\ref{fig:evol} shows the evolution of the continuum intensity
and the magnetic field in a $2.9\times2.2$~Mm subdomain of the
simulation, over approximately $7$ minutes. Shown are maps at the
original resolution of the simulation (left-hand side) and at the
spatial resolution of Hinode (right-hand side). For the original
resolution, we show the vertical component of the magnetic field
from the simulations (right column on the left-hand side). The
longitudinal apparent flux density \citep{Lites:etal:2008}
retrieved from synthetic Stokes profiles is shown at the Hinode
resolution (right column on the right-hand side). During the
period shown, three bright points appear near the coordinates
$[2\arcsec,1.5\arcsec]$, $[0.5\arcsec,1\arcsec]$ and
$[3.5\arcsec,1\arcsec]$, in the regions outlined by yellow
squares. We refer to them here as cases I, II and III,
respectively. They are identified in the top left frame.

\subsection{Horizontal flows}
\label{sec:hor_flows}

In all three cases, magnetic field is advected by the flow to the
junction of multiple granules, where it is confined and
concentrated. The evolution of magnetic concentrations is examined
more closely in Fig.~\ref{fig:evol_zoom}, where the regions
outlined by yellow squares in Fig.~\ref{fig:evol}, are enlarged.
Horizontal velocities, at approximately $80$~km above
$\langle\tau\rangle=1$, are represented by arrows in continuum
maps and magnetograms. Regions with a strong vertical component of
vorticity are also outlined.

Vortex flows around strong downflows are quite common in
simulations of granulation \citep{Nordlund:1986}. They are formed
at the vertices between multiple granules, where flows converge
and angular velocities with respect to the center of downflow
increase due to angular momentum conservation. The lifetime of
vortex flows depends on the dynamical behavior of the neighboring
granules. In a $30$~min run that we examined, they last from less
then $90$~s to approximately $10$~min. The ones shown in
Fig.~\ref{fig:evol_zoom} are short-lived. In case II, the swirling
motion persists for at least 150 s, until the shape of the
neighboring granules is changed. The formed flux concentration is
then squeezed between two granules and stretched into a
flux-sheet-like feature. Later on, as the granules evolve, the
magnetic feature is advected to the left and trapped again in the
junction between newly formed granules. In case III, a strong
preexisting flux sheet is carried by the flow toward the junction
of the granules. There, the field is caught in a vortex flow that
lasts for at least $100$~s. The flow becomes disturbed and then
starts to swirl in the opposite direction. The vortex axes are
inclined, in both cases II and III, with respect to the vertical
direction. Also in both cases, there is a spatial and temporal
coincidence between the existence of vortex flows and the
formation of high speed downflows near the surface layer.

The vortices shown in Fig.~\ref{fig:evol_zoom} are less then
$0.25\arcsec$ in diameter, much smaller then the ones found in
observations \citep{Bonet:etal:2008}. The size of the observed
vortex flows is of order $1$\arcsec and the average lifetime is
around 5 min. They seem to outline supergranulation and
mesogranulation cells. As the authors of the observational study
suggest, most vortices could have been missed because of the
limited spatial resolution (Swedish Solar Telescope) and the
method used (tracking the motion of bright points in G band). A
short-lived vortex flow preceding formation of a network bright
point has been observed by \citet{Roudier:etal:1997}.

No vortex is detected during the formation of flux concentration
in the case I. In this case, a strong downflow, and later upflow
develops inside the flux concentration, whose shape and contrast
in the intensity map is greatly affected by the convective
motions. Spatial and temporal fluctuations across the magnetic
feature make it difficult to define its center. The bright points,
visible in continuum maps, form where fragments of flux
concentrations become narrow so that lateral heating becomes
important and in the regions with higher field strength, i.e.
higher evacuation.

During their further evolution after the intensification phase ($t
> 420$~s), the flux concentrations in cases I, II and III have
different fates. In case II, it meets a magnetic feature of
opposite polarity and vanishes in $\sim 3$~minutes. The feature
formed in case I is gradually stretched by the flow and fragmented
after approximately $9$~minutes. A few minutes after that, the
larger fragment encounters the opposite-polarity flux
concentration from case III. The cancellation continues for $\sim
12$~minutes until both features disappear. The smaller fragment of
the feature from case I is caught in a granular vortex flow
approximately $5$ minutes later which leads to further
intensification.

\begin{SCfigure*}
    \centering
    \includegraphics[width=0.75\textwidth]{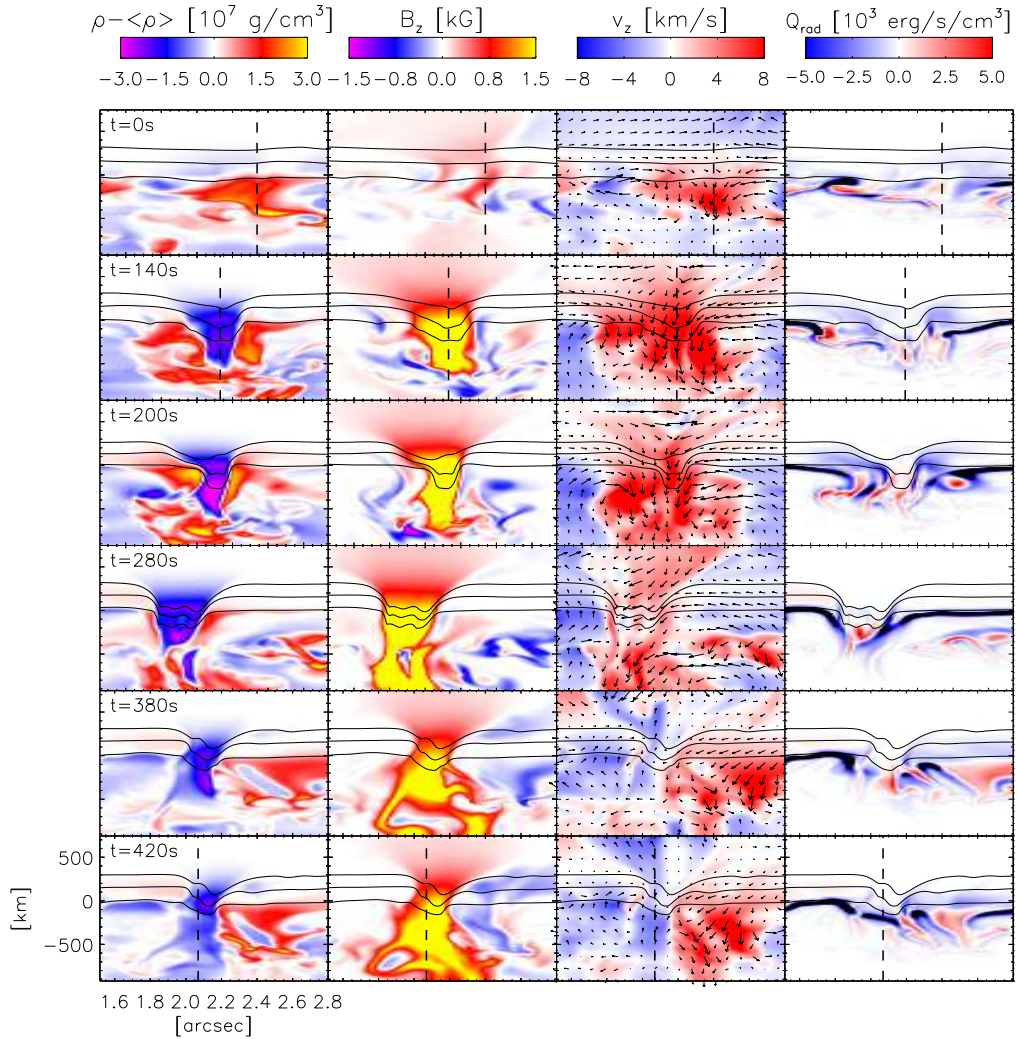}
    \caption{Time sequence of vertical cuts through the magnetic concentration corresponding to case I
    (positions of the cuts are marked by yellow horizontal lines in right-hand images in
    Fig.~\ref{fig:evol}). The plotted cuts correspond to the same instances of time as sampled in
    Fig.~\ref{fig:evol}. From left to right: density excess with respect to the the mean density at every
    height, vertical component of magnetic field, vertical component of velocity (overplotted arrows show the
    components of the velocity field in the x-z plane), and radiative heating rate.  Horizontal solid lines follow the levels of $\log~\tau = 0, -1$ and $-2$.
    Vertical dashed lines mark the position of the height profiles given
    in Fig.~\ref{fig:1dh}.}
    \label{fig:cut}
\end{SCfigure*}

\begin{figure}
   \centering
\includegraphics[width=0.85\hsize,angle=90]{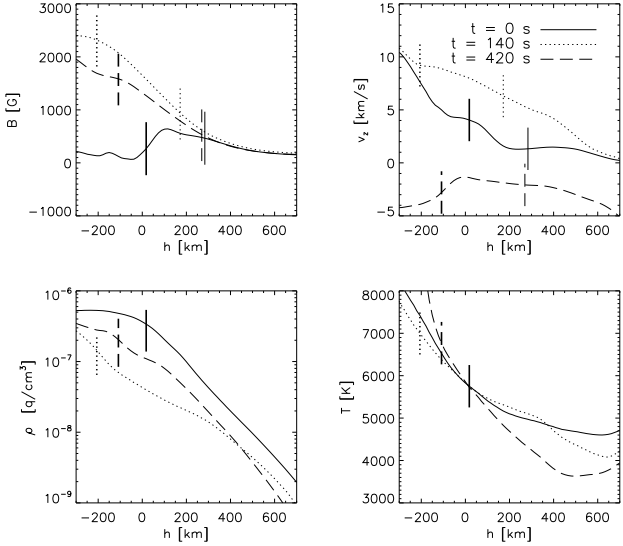}
   \caption{Height profiles of various quantities in the flux concentration corresponding to the case I (magnetic field, vertical component of
   velocity, density and temperature) at the positions marked by dashed lines in
   Fig.~\ref{fig:cut} at different time instances. Thick vertical lines mark the positions of the optical depth $\tau_{500} = 1$ for each atmosphere.
    Thin vertical lines in the top two panels mark the positions of the height of formation of Fe~I~630.25~nm line (see text).}
   \label{fig:1dh}
\end{figure}

\subsection{Field intensification}

The evolution of the magnetic structures formed during the studied
time interval can be followed in more detail by taking vertical
slices through the regions marked by horizontal lines in
Fig.~\ref{fig:evol_zoom}. We will examine three cases.

\subsubsection{Case I}

Vertical cuts through the regions corresponding to this case are
given in Fig.~\ref{fig:cut}. Shown are the density excess
(relative to the mean density at each geometrical height), the
vertical components of magnetic field and velocity, and the
radiative heating rate. Cuts at the instant $t=0$ show the
fluctuation in density outlining the granulation structure, with
less dense, hotter granules and a cooler intergranular lane, where
magnetic field of a few hundred G is accumulated and a downflow of
a few km/s is present. The fluctuation in radiative heating rate
is highest in the thin layer near optical depth unity, where
upflowing granular material cools most strongly. The regions along
the intergranular lane seems to undergo cooling in the layers
above $\tau=1$. In the next instant, $140$~s later, a strong
magnetic concentration has already formed. Variations in magnetic
field strength at a given geometrical height are accompanied by
variations in the gas pressure. Regions with increased field
strength have lower density, which, in turn, shifts the level of
optical depth unity downwards relative to neighboring regions with
weaker field. The depression of the visible surface inside the
magnetic structure gives rise to radiative heating through the
sidewalls from the hot neighboring material. The downflow velocity
is significantly increased from $t=0$ to $t=140$~s, becoming
supersonic in the layer between 100 km above and 350 km below the
surface. By $t=280$~s it is greatly diminished again inside the
flux concentration. The diameter and the shape of the flux
concentration continuously evolves and does not reach a stationary
state. At $t=280$~s strong radiative heating is present inside the
concentration. This is also the moment when continuum intensity is
the highest as shown later (Fig.~\ref{fig:evol_red}).

In the last two sets of snapshots, shown in Fig.~\ref{fig:cut}, an
upflow is visible inside the flux concentration, extending almost
from the bottom of the simulation domain. It reaches supersonic
velocities in the upper layers of the photosphere. Its direction
is inclined with respect to the vertical. The upflowing material
inside the concentration exhibits strong cooling at the surface.
As the material refills the region inside the flux concentration,
the level of optical depth unity shifts upwards and the magnetic
field strength decreases. However, the upflow does not lead to a
complete dispersal of the field, as described in the last
paragraph of Sec.~\ref{sec:hor_flows}.

The evolution of this flux concentration is further illustrated by
the vertical profiles given in Fig.~\ref{fig:1dh}. Three phases
before and after the field intensification are shown. The
horizontal positions of the profiles are marked by vertical dashed
lines in Fig.~\ref{fig:cut}. The profiles of magnetic field
strength, vertical component of velocity, density and temperature
for each phase are shown, together with the corresponding level of
optical depth unity, marked by thick vertical lines. The level
$h=0$ corresponds to the level of the mean optical depth unity
$\langle\tau_{500}\rangle=1$ for the whole snapshot. Thin vertical
lines in the upper plots mark the position of the average height
of formation of the Fe I 630.25 nm line, which is defined by
calculating the centroid of the contribution function (CF) for
line intensity depression \citep{Solanki:Bruls:1994}, at the
wavelength of the line core. The definition gives only a rough
estimate of the heights sampled, since the lines are formed over a
large portion of the photosphere and the CFs are asymmetric in
wavelength due to the presence of high velocity gradients.

The figure shows that, in the interval from $t=0$~s to $t=140$~s,
the downflow extends to deeper layers and increases in amplitude,
reaching 10 km/s at 200 km below the surface. The magnetic field
strength increases from a few hundred to more than 2000 G at the
level of $\tau_{500}=1$. The significant reduction in density
results in a shift of 200 km in the optical depth unity level. As
a result of radiative heating the temperature gradient is flatter
and the temperature is around 400 K higher at $h=0$. The
temperature difference at equal optical depth is even larger,
exceeding 1000 K at $\tau_{500}=1$. The estimated height of
formation of Fe I 6302 shifts from roughly 300 km to 150 km above
the surface when magnetic field concentration forms, which is in
agreement with \citet{Khomenko:Collados:2007}. This effect
contributes to the increase of observed magnetic field strength in
the first two phases.

In the third phase, the upflow leads to a density enhancement and
a weakening of the magnetic field by a few hundred Gauss in the
lower photosphere. The level of $\tau_{500}=1$ shifts upward by
100 km. No discontinuity is visible in the vertical component of
velocity, in contrast to the event studied by
\citet{GrossmannDoerth:etal:1998}.

\begin{SCfigure*}
    \centering
    \includegraphics[width=0.75\textwidth]{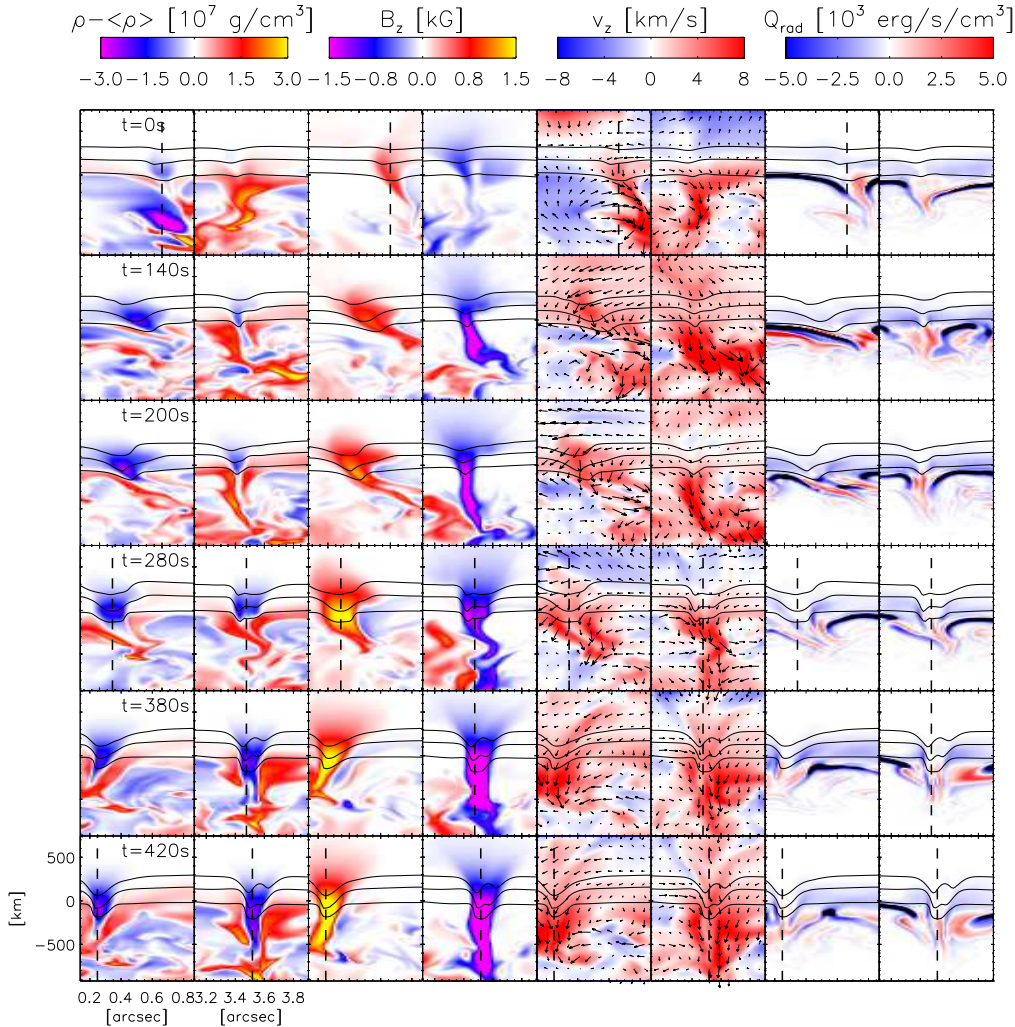}
    \caption{The same as Fig.~\ref{fig:cut} but for cases II and
    III. Each frame is subdivided into a left part displaying case
    II and a right part showing case III.
    Vertical lines mark the positions of the height profiles given
    in Fig.~\ref{fig:1dh_f2} and Fig.~\ref{fig:1dh_f3}.}
    \label{fig:cut_f23}
\end{SCfigure*}

\begin{figure}
   \centering
  \includegraphics[width=0.85\hsize,angle=90]{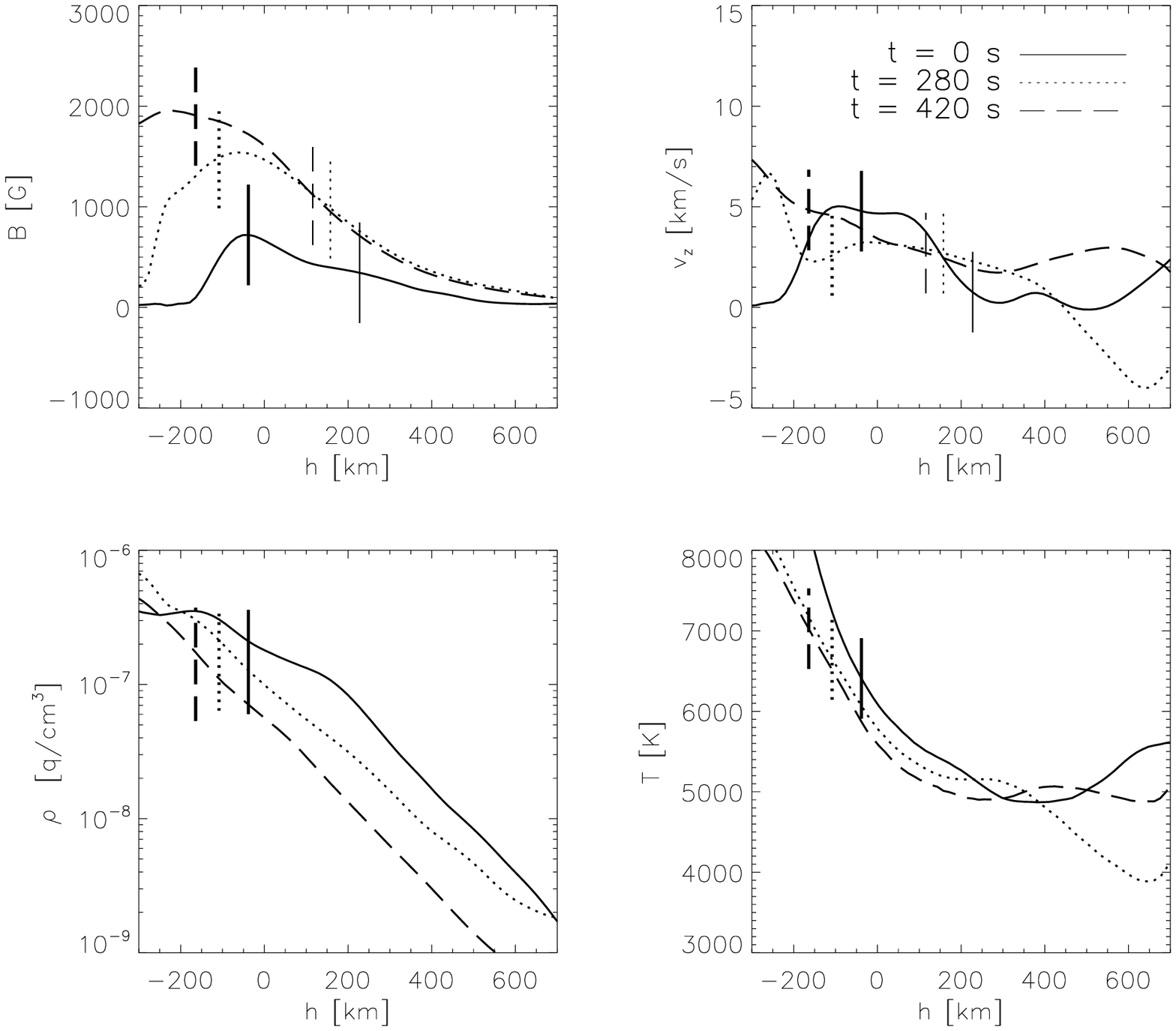}
   \caption{The same as Fig.~\ref{fig:1dh}, but for case II.}
   \label{fig:1dh_f2}
\end{figure}

\begin{figure}
   \centering
\includegraphics[width=0.85\hsize,angle=90]{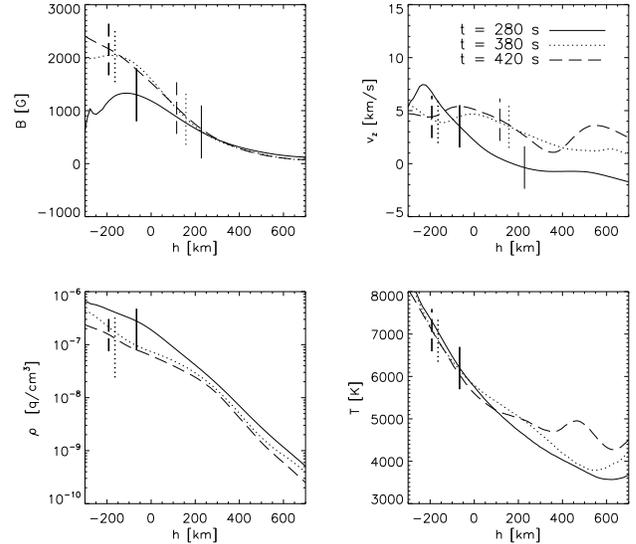}
   \caption{The same as Fig.~\ref{fig:1dh}, but for case III.}
   \label{fig:1dh_f3}
\end{figure}

\subsubsection{Cases II and III}

Vertical cuts through the regions corresponding to these cases are
shown in Fig.~\ref{fig:cut_f23}. The positions of the cuts are
marked by horizontal lines in Fig.~\ref{fig:evol_zoom}. The
parameters shown are the same as in Fig.~\ref{fig:cut}.

At $t=0$, weak magnetic structures in the intergranular lanes are
already present in both cases. Strong downflows along the magnetic
field lines persist over the entire time interval shown. They
reach supersonic velocities in some instances, but only in the
layers below $\tau =1$. In case II, the downflow is strongest at
$t=140$~s, just before the vortex flow becomes deformed owing to
evolution of the neighboring granule. Then it gets somewhat
reduced at t=280s, as the structure is squeezed between the
granules. Later, at $t=380$~s, a strong downflow is present again
in the newly formed granular junction, to which the flux
concentration is carried and where it is confined again. The
rightmost columns show strong cooling of the layers above the
level of $\tau_{500}=1$ in both cases. At $t=420$~s, the interior
of the flux concentrations reaches radiative equilibrium. At that
moment, the flux concentrations in both cases are almost vertical,
with diameters of approximately $0.1\arcsec$ at $\tau_{500}=1$.

Figures~\ref{fig:1dh_f2} and \ref{fig:1dh_f3} show height profiles
along the positions marked by the dashed lines in
Fig.~\ref{fig:cut_f23}. Since the flux concentrations are
inclined, the choice of curved lines, along the axes of magnetic
structures, would be more logical here. However, we are interested
in what would be observed at the disc center and the observables
(i.e. the Stokes profiles) depend on the vertical profiles of the
physical parameters which are shown here. In the case II, we
choose to show profiles at instants $t=0$, $280$ and $t=420$~s
(solid, dotted, dashed respectively). For the case III, profiles
at the last three instants from Fig.~\ref{fig:cut_f23} are shown.
Both cases exhibit a similar behavior: there is a significant
evacuation of the magnetic concentration due to the strong
downflow and an increase of the magnetic field strength from a few
hundred G to kG values in the process. There is a persistent
dowflow that reaches 5 km/s (in the vertical direction) at
$\tau_{500}=1$ and a decrease in density with a corresponding
shift in the height of $\tau_{500}=1$. In both cases, the shift in
$\tau_{500}=1$ reaches 150 km and the magnetic field strength
increases to 2000 G at that depth. The height of formation of the
Fe I 630.25 nm line shifts to deeper layers as the magnetic field
strength increases. In the last instant, the Fe I 630.25 nm line
probes layers with kG fields and vertical velocity of 5 km/s, in
both cases.

\begin{figure*}
    \centering
    \includegraphics[width=0.6\hsize,angle=90]{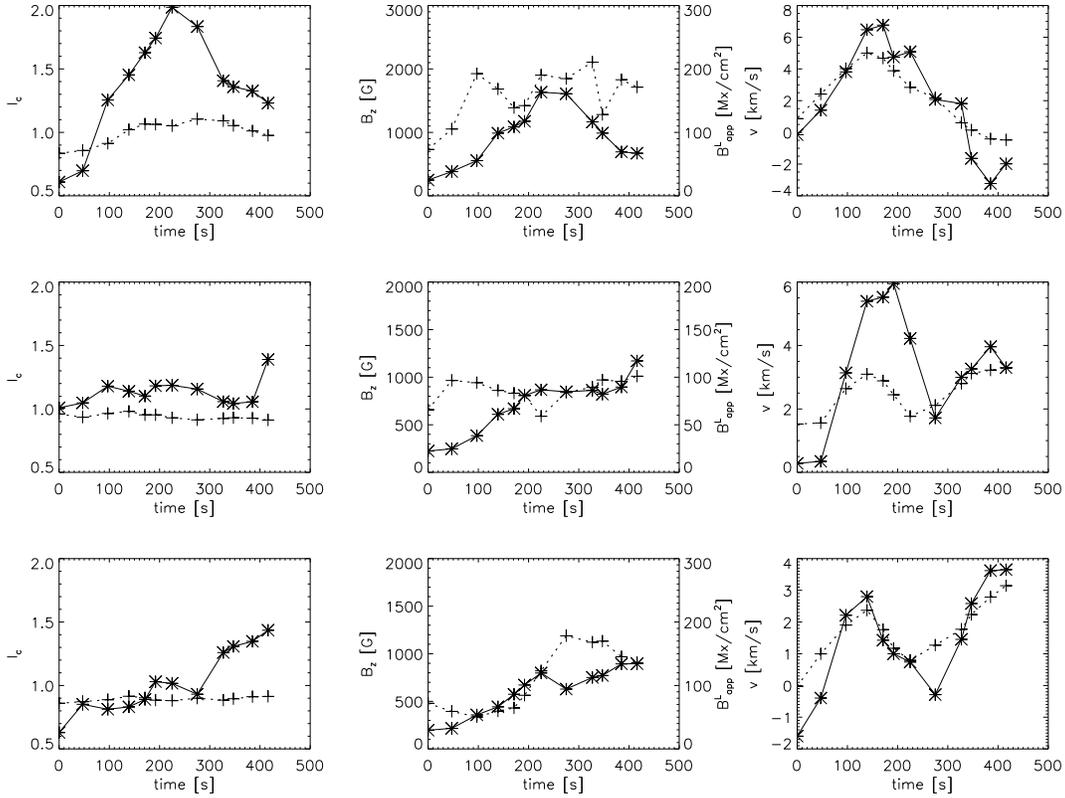}
    \caption{Temporal change of normalized continuum intensity (left), vertical component of magnetic field (middle) and vertical component of velocity
    (right), for cases I (top row), II (middle row) and III (bottom row) at original (stars/solid line) and Hinode resolution (crosses/dashed line).
    Vertical component of magnetic field and velocity at log~$\tau_{500} = -2$ (for original resolution) are each plotted together with
    the longitudinal apparent magnetic flux density (note the separate scale employed for B$^{L}_{app}$ in the middle column) and velocity retrieved from Stokes V profiles (Hinode resolution), respectively. }
    \label{fig:evol_red}
\end{figure*}

\begin{figure}
    \centering
    \includegraphics[width=0.4\hsize,angle=90]{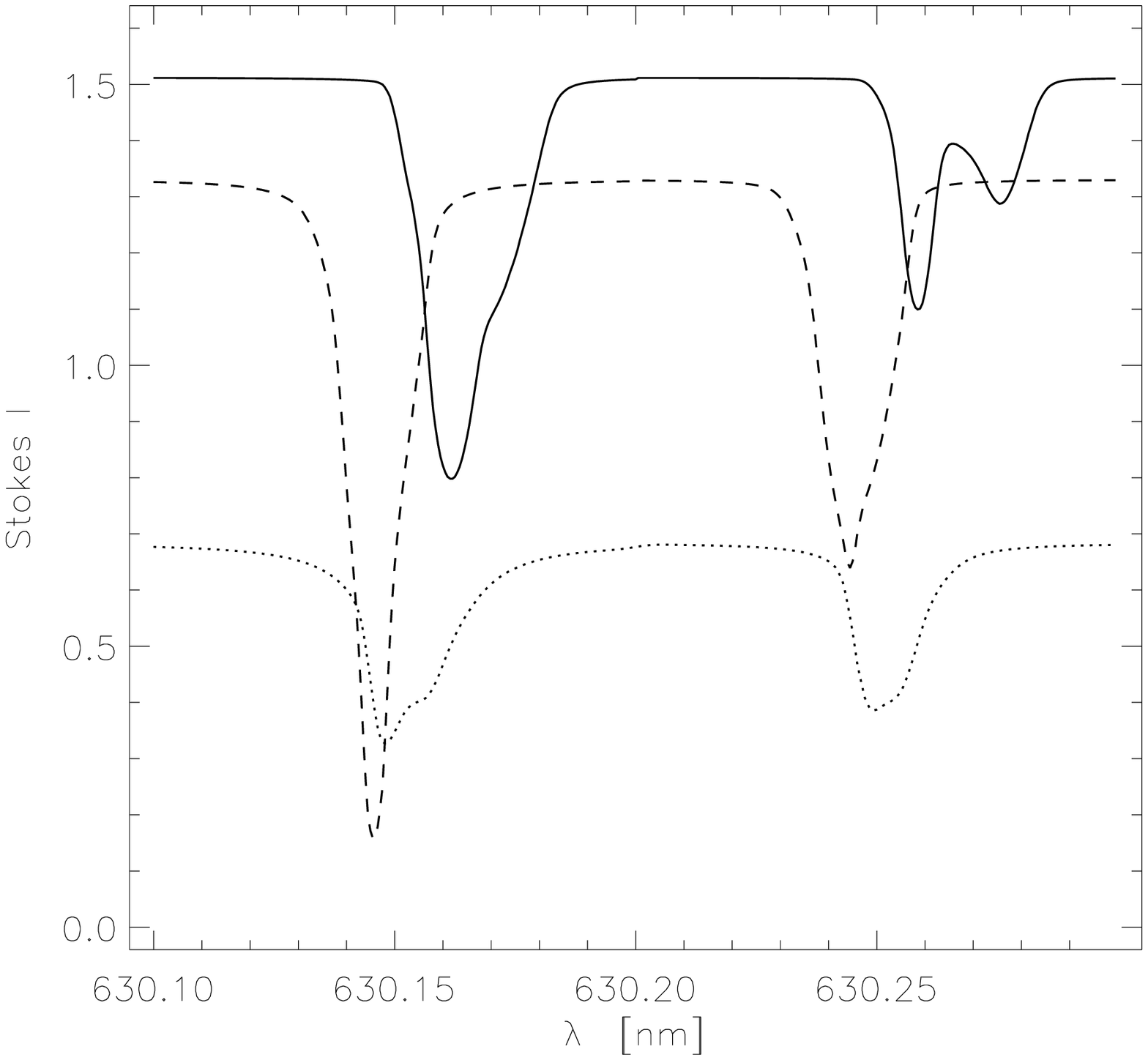}
    \includegraphics[width=0.4\hsize,angle=90]{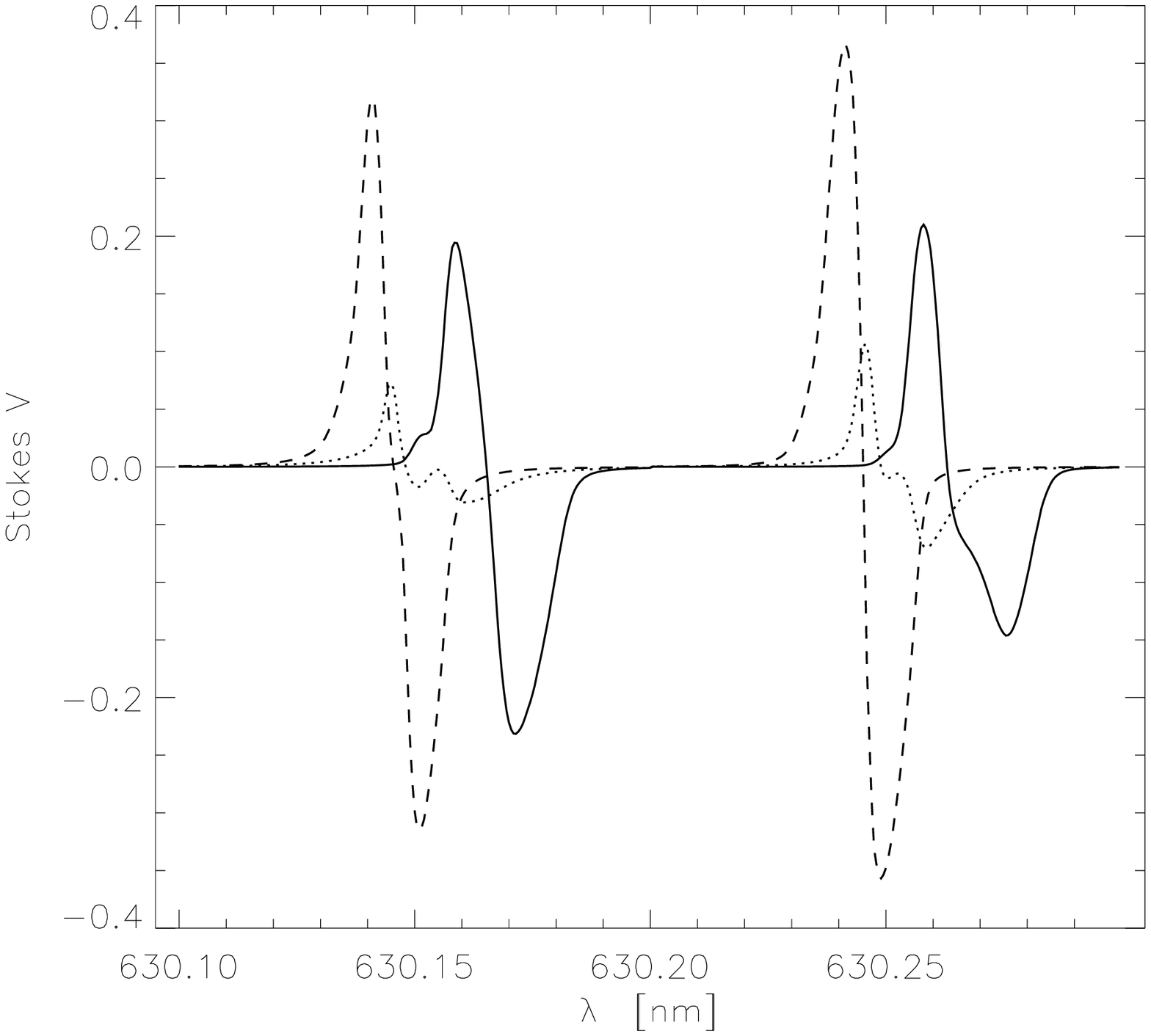}
    \includegraphics[width=0.4\hsize,angle=90]{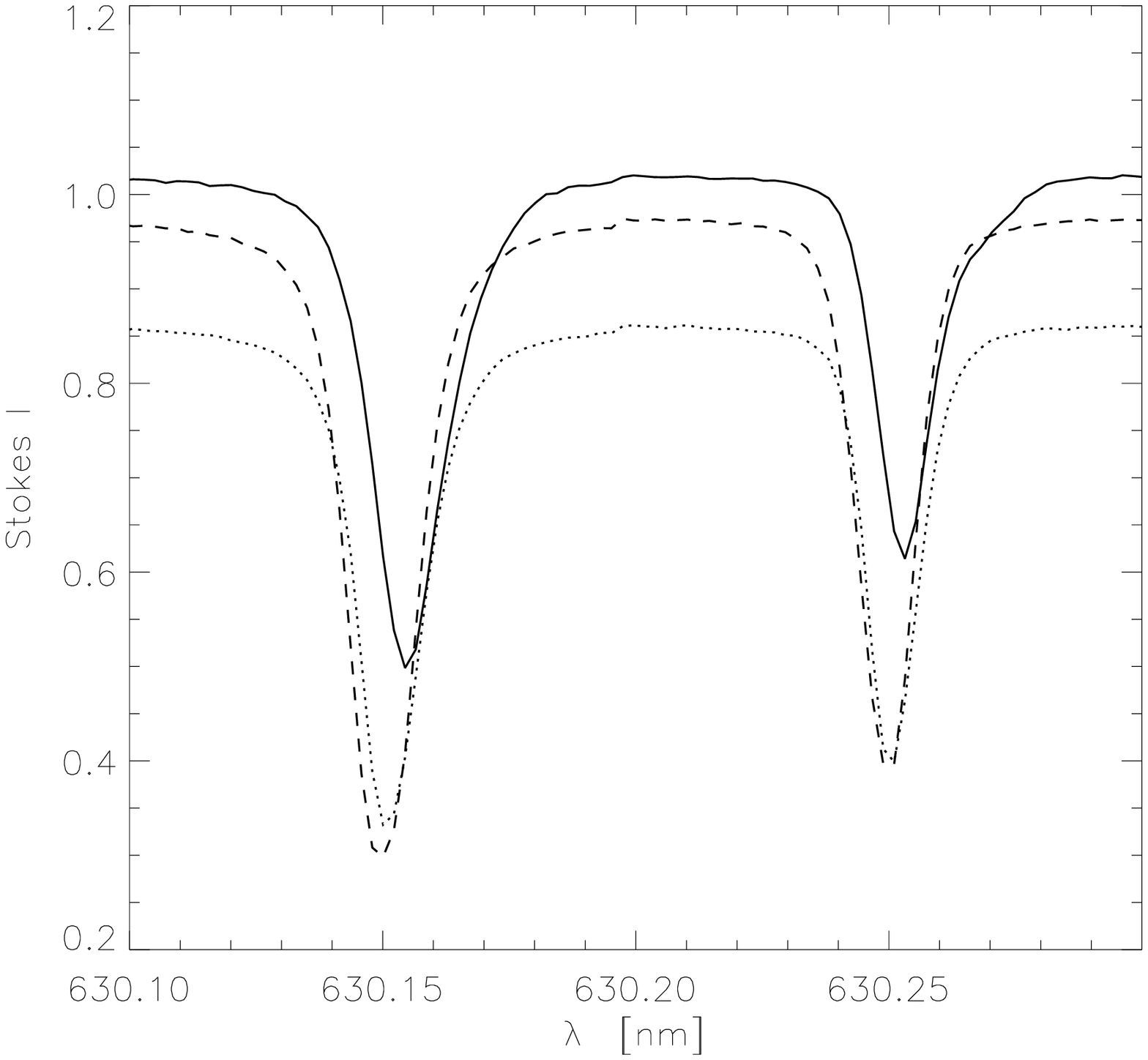}
    \includegraphics[width=0.4\hsize,angle=90]{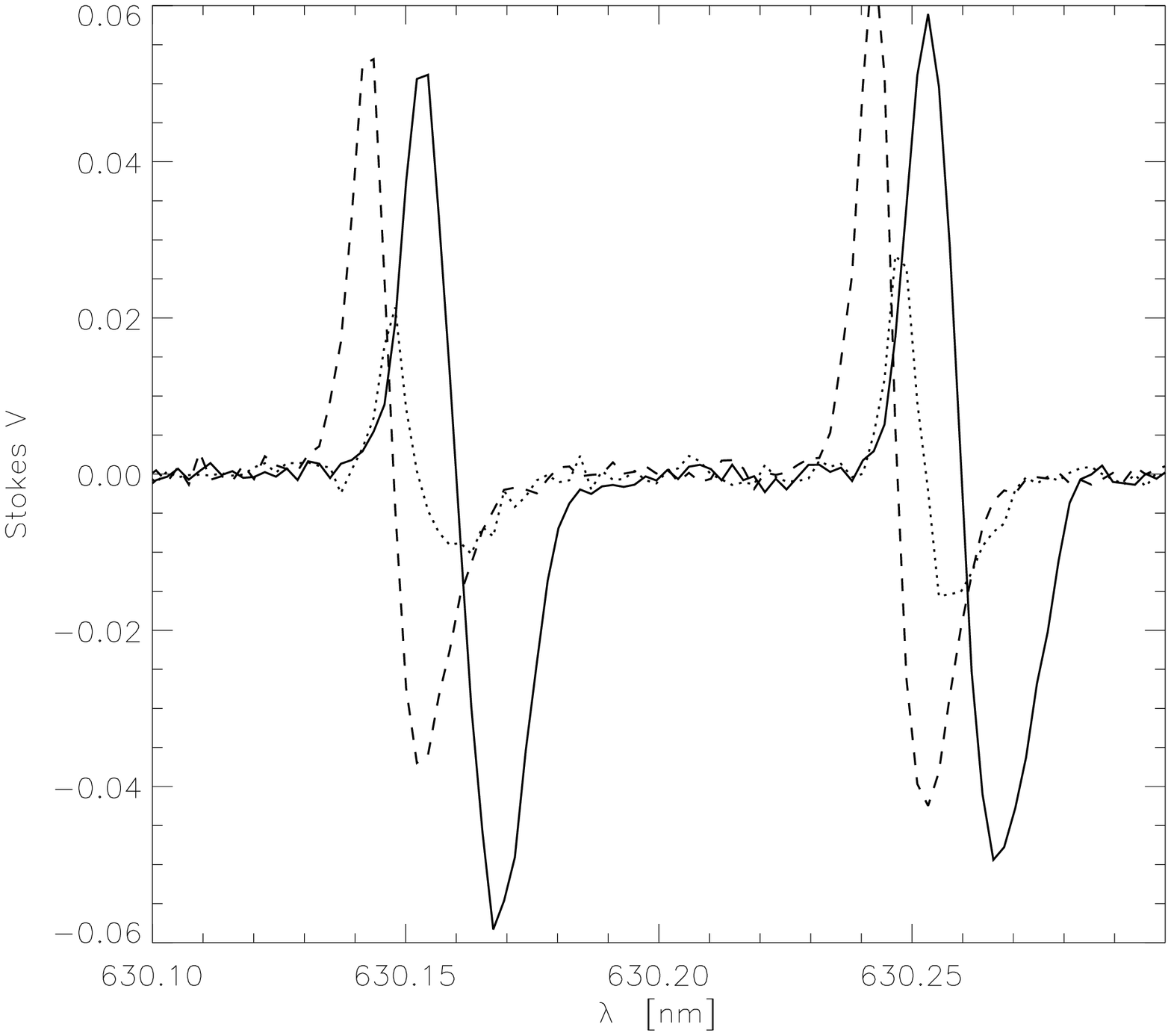}
    \caption{Synthetic Stokes I (left) and V (right) profiles at original (upper row) and Hinode resolution
    (lower row) for the times $t=0,140$ and $420$ s marked by dotted, solid and dashed,
    respectively. The intensity is normalized to the mean continuum intensity in
    the full snapshot domain.}
    \label{fig:stokes}
\end{figure}

\subsection{Comparison with synthetic Hinode observations}

The intensity contrast is considerably reduced at Hinode spatial
resolution, as can be seen in Fig.~\ref{fig:evol} (note that the
color scales are adjusted). The maps of longitudinal apparent
magnetic flux density are shown for comparison with the
magnetograms at the original resolution. Fine structure in both
maps, continuum and magnetograms, is lost due to smearing. The
bright points that correspond to cases II and III are barely
discernible.

\subsubsection{Case I}

In order to compare case I with the results of
\citet{Shimizu:etal:2008}, we calculate the parameter they refer
to as \textit{signal excess} (SE). It is defined as the Stokes $V$
profile integrated over the spectral range of $250-400$~m\AA~
redward from the nominal line center. This wavelength range
corresponds to dowflows of $7-14$ km/s, so that high values of SE
indicate the presence of hight-speed downflows. The red contour in
the right-hand side of Fig.~\ref{fig:evol} outlines a region with
SE of $0.01$~pm, which is a factor of 10 higher than in the
surroundings. It fits the location and time of the strong
downflows visible at the original resolution (left hand-side of
Fig.~\ref{fig:evol}). Both, synthesized (here) and observed
\citep{Shimizu:etal:2008} SE occur simultaneously with the
appearance of the bright point and the intensification of the
magnetic field. However, the event described by
\citet{Shimizu:etal:2008} (Fig. 8) lasts at least $6$ minutes,
while in our simulations it is present for less than $4$ minutes.

Figure~\ref{fig:evol_red} compares the observable parameters with
the corresponding values at original resolution. The locations
chosen for the plots are marked by yellow crosses in
Fig.~\ref{fig:evol}. The pixels are selected such that regions of
downflow and upflow are covered, as well as the evolution of the
bright point. The plots show the temporal change of normalized
intensity, longitudinal apparent flux density
\citep{Lites:etal:2008}, and zero-crossing velocity retrieved from
Stokes $V$ profiles of the Fe I 630.25 nm line, so that a direct
comparison with the results of \citet{Nagata:etal:2008} can be
made. Overplotted (stars/solid lines) are the normalized intensity
at the original resolution, the vertical component of magnetic
field and velocity at $\log\tau_{500}=-2$.

The comparison shows that the correlation between the parameters
at original and reduced resolution is not so good. This is a
consequence of different effects. Firstly, since the formation
region of the Fe I 630 nm lines extends over a large part of the
atmosphere, there is smearing along the line of sight and the
resulting profiles carry information from different layers with
different values of the physical parameters. Secondly, the
smearing due to spatial resolution affects every pixel differently
depending on its neighboring pixels. The profiles could be more
asymmetric and broader due to averaging over inhomogeneous regions
and the amplitude of Stokes $V$ can be affected.
Figure~\ref{fig:stokes} shows examples of Stokes $I$ and $V$
profile at original and reduced resolution at  $t=0,~140$ and
$420$~s. The profiles show stronger asymmetries and higher Doppler
shifts at original resolution. The amplitudes of Stokes $V$
profiles at Hinode resolution are reduced by a factor of 10. They
are of the same order of magnitude as the observed ones (Fig. 2 of
\citet{Nagata:etal:2008}). Both, synthesized and observed Stokes
$I$ profiles show an extended red wing at the phase with strong
downflow.

Figure~\ref{fig:evol_red} reveals that intensity contrast is
significantly reduced by spatial smearing. Thus at $t=230$~s the
contrast of the bright point drops from 100\% at original
resolution to below 30\% at the reduced resolution. The velocity
determined from Stokes V profiles is at most times somewhat lower
than the velocities at $\log\tau_{500}=-2$. However, the general
trends of the parameters are preserved. For both, simulation and
observation \citep{Nagata:etal:2008}, the intensity reaches its
maximum approximately 100 s later than velocity. This is
consistent with the idea that the enhanced brightness results from
the evacuation due to the downflow. When the velocity changes
sign, the brightness starts to fade. The maximum brightness of the
simulated and observed bright points are similar after degrading
the simulated intensity to Hinode resolution, while the velocities
derived from the simulations are consistent, but somewhat smaller.
They reach approximately $5$~km/s (downflow) and $0.5$~km/s
(upflow), while values up to $6$~km/s and $2$~km/s, respectively,
are observed. The magnetic field strength obtained by
\citet{Nagata:etal:2008}, by inversions, increases from a few
hundred to $2000$~G in approximately $200$~s, which is in
agreement with simulations.

\subsubsection{Cases II and III}

The middle and bottom rows in Figure~\ref{fig:evol_red} show the
temporal evolution for cases II and III, respectively. We find a
simultaneous increase in all parameters (continuum intensity,
apparent flux density and zero-crossing velocity). In both cases,
the downflow is suppressed at $t=280$~s possible because of the
evolution of the surrounding granules. The velocity at
$\log\tau_{500}=-2$ reaches 6 and 4 km/s for case II and III,
respectively. As the magnetic field strength increases, the
intensity follows and, in both cases, the intensity contrast
reaches 50\% with respect to the mean at original resolution.

The influence of spatial smearing is on the whole similar,
although a few differences relative to case I are noteworthy.
Firstly, due to the small size of the features, the correlation of
the magnetic and brightness signals is largely destroyed. Although
the magnetic signals are comparable to the values in case I, the
brightness is lower, so that the features are inconspicuous in
continuum images. The increase in magnetic flux density is
followed by just a small fluctuation in intensity. Secondly,
although the magnetic field strengths and diameters of the
features are comparable for cases II and III at original
resolution, the case II shows lower magnetic flux density after
convolution. Thirdly, the signature of the strong dowflow in case
II is lost due to spatial smearing.

\section{Summary}

Our case study of three examples of magnetic field intensification
has shown that in all three cases, the field is advected to the
junction of several granules. There, it is confined by converging
granular flows, which, in two cases, form a vortex. Owing to the
presence of the magnetic field, the thermal effect (radiative
cooling) induces evacuation of the flux concentration. The
evacuation leads to a downward shift of the optical depth scale
within the flux concentrations. The shift is smaller for the
smaller features due to the lateral radiative heating, which
inhibits further evacuation \citep{Venkatakrishnan:1986}. As a
result, the magnetic field at $\tau=1$, in the smaller features,
is weaker than in the case of the feature with more flux. This is
in accordance with recent numerical \citep{Cheung:etal:2007} and
observational \citep{Rueedi:etal:1992,Solanki:etal:1996} results.

During the evacuation, the dowflow velocities reach maximum values
of 5-10 km/s at $\tau=1$. In the case of the largest feature, the
downflow extends from the upper boundary of the simulation domain
and becomes supersonic in the lower photosphere. The magnetic
features formed have diameters of 0.1-0.2~\arcsec. In the case of
the biggest feature, a supersonic upflow develops approximately
$200$~s after the formation of the flux concentration. The upflow
does not lead to a complete dispersal of the field, but the
feature persists until 9 minutes later when it undergoes a
fragmentation. The disappearance of the features in all three
cases occurs when they meet opposite polarity features, between 3
and 20 minutes after their formation.

We also show what happens with the observables when the effects of
smearing to observational spatial resolution is taken into
account. An important result is that, in the case of small
features, Hinode/SP would miss the bright point formation and, in
some cases, also the high velocity downflows that develop in the
process. On the other hand, the signatures of the evolution of
large features are detectable even after the spatial smearing. We
show that this case can be quantitatively compared with Hinode/SP
observation \citep{Nagata:etal:2008,Shimizu:etal:2008} and
exhibits a very similar evolution. This suggests that the magnetic
field intensification process in the MURaM simulations is a
faithful description of the process taking place on the Sun.
Furthermore, our study indicates that the analysis and
interpretation of the observations in terms of the convective
intensification process is well-founded.

\begin{acknowledgements}
Hinode is a Japanese mission developed and launched by ISAS/JAXA,
with NAOJ as domestic partner and NASA and STFC (UK) as
international partners. It is operated by these agencies in
co-operation with ESA and NSC (Norway). We thank R. Cameron for
valuable suggestions. This work was partially supported by WCU
grant No. R:31-0016 funded by the Korean Ministry of Education,
Science and Technology.
\end{acknowledgements}

\bibliographystyle{aa}

\end{document}